\documentclass[a4paper,12pt]{article}
\usepackage[english]{babel}
\usepackage{graphicx}
\usepackage{amssymb}
\usepackage{amsmath}
\usepackage{geometry}
\usepackage{hyperref}
\hypersetup{
colorlinks=true,
linkcolor=blue,
citecolor=blue,
urlcolor=black
}

\begin{document}

\newcommand{\eqd}{\stackrel{d}{=}}
\renewcommand{\r}{\mathbb R}
\renewcommand{\Pr}{\mathbb P}

\renewcommand{\refname}{References}

\title{\vspace{-1.8cm}
A functional approach to estimation of the parameters of generalized negative binomial and gamma distributions}

\author{
A.\,K.~Gorshenin\textsuperscript{1},
V.\,Yu.~Korolev\textsuperscript{2}}

\date{}

\maketitle

\footnotetext[1]{Institute of Informatics Problems, Federal Research
Center ''Computer Science and Control'' of Russian Academy of
Sciences, Russia; \url{agorshenin@frccsc.ru}}

\footnotetext[2]{Faculty of Computational Mathematics and
Cybernetics, Lomonosov Moscow State University, Russia; Institute of
Informatics Problems, Federal Research Center ``Computer Science and
Control'' of Russian Academy of Sciences, Russia; Hangzhou Dianzi
University, China; \url{vkorolev@cs.msu.su}}

\maketitle

\begin{abstract}
The generalized negative binomial distribution (GNB) is a new
flexible family of discrete distributions that are mixed Poisson
laws with the mixing generalized gamma (GG) distributions. This
family of discrete distributions is very wide and embraces Poisson
distributions, negative binomial distributions, Sichel
distributions, Weibull--Poisson distributions and many other types
of distributions supplying descriptive statistics with many flexible
models. These distributions seem to be very promising for the
statistical description of many real phenomena. GG distributions are widely applied in signal and image processing and other practical problems. The statistical estimation of the parameters of GNB and GG distributions is quite complicated. To find estimates,
the methods of moments or maximum likelihood can be used as well as
two-stage grid EM-algorithms. The paper presents a methodology based
on the search for the best distribution using the minimization of
$\ell^p$-distances and $L^p$-metrics for GNB and GG distributions,
respectively. This approach, first, allows to obtain parameter
estimates without using grid methods and solving systems of
nonlinear equations and, second, yields not point estimates as the
methods of moments or maximum likelihood do, but the estimate for
the density function. In other words, within this approach the set
of decisions is not a Euclidean space, but a functional space.
\end{abstract}

\section{Introduction}
The generalized negative binomial distribution (GNB) is a new
flexible family of discrete distributions that are mixed Poisson
laws with the mixing generalized gamma (GG) distributions. The GNB
distributions were introduced and studied
in~\cite{Korolev2017} under the name of GG mixed Poisson
distributions. This family of discrete distributions is very wide
and embraces Poisson distributions (as limit points corresponding to
a degenerate mixing distribution), negative binomial (Polya)
distributions including geometric distributions (corresponding to
the gamma mixing distribution, see~\cite{Greenwood1920}), Sichel
distributions (corresponding to the inverse gamma mixing
distributions, see~\cite{Sichel1971}), Weibull--Poisson
distributions (corresponding to the Weibull mixing distributions,
see~\cite{Korolev2016}) and many other types supplying
descriptive statistics with many flexible models. These
distributions seem to be very promising for the statistical
description of many real phenomena being very convenient and almost
universal models. It is quite natural to expect that, having
introduced one more free parameter into the pure negative binomial
model, namely, the power parameter in the exponent of the original
gamma mixing distribution, instead of the negative binomial model
one might obtain a more flexible GNB model that provides even better
fit with the statistical data. For example, GNB distributions can be
successfully applied to modeling statistical regularities in
duration of specific periods in data.

The GG distributions are proposed in order to have a flexible
Bayesian model with a mixing (prior) distribution which is
``responsible'' for the description of statistical regularities of
the manifestation of external stochastic factors. The class of
GG distributions was first described as a unitary family in $1962$
by E. Stacy~\cite{Stacy1962}. The family of GG distributions
contains practically all the most popular absolutely continuous
distributions concentrated on the non-negative half-line including
Weibull and gamma distributions.

GG distributions are widely applied in many practical problems.
There are dozens of papers dealing with the application of
GG distributions as models of regularities observed in practice.
As an example, the following research areas involving
models based on GG distributions can be mentioned:
\begin{itemize}
\item climatic and hydrological problems: drop size distributions~\cite{Maur2001},
drought data~\cite{Nadarajah2007},  phenomena in warm clouds~\cite{Xie2009};
\item synthetic-aperture radar (SAR) image processing and various applications:
distribution for the real and imaginary parts of the complex SAR backscattered signal~\cite{Li2010},
flexible model for the SAR images with different land-cover typologies~\cite{Li2011},
statistical modeling of SAR images~\cite{Qin2015,Sportouche2017};
\item astrophysical problems, for example, new galaxy luminosity functions~\cite{Zaninetti2010};
\item speech signal processing: parametric characterization of speech spectra~\cite{Shin2005},
modelling speech samples~\cite{Li2008}, real-time implementations of algorithms~\cite{Song2008}.
\end{itemize}

Apparently, the popularity of GG distributions is due to that most
of them can serve as adequate asymptotic approximations, since all
the representatives of the class of GG distributions listed above
appear as limit laws in various limit theorems of probability theory
in rather simple limit schemes.

The problem of statistical estimation of the parameters
of GNB and GG distributions (for example, the search for maximum
likelihood (ML) estimates) is quite complicated. To find the
estimates of the parameters, the method of moments or ML
method~\cite{Huang2006} for the GG distribution as well as the
two-stage grid EM-algorithm for the GNB, can be used. It should be
noted that the implementations of the methods of moments and ML
method  for GG distribution are difficult computational tasks,
moreover, the efficiency depends on the sample size (ML method is
better for large volumes).

The paper presents a methodology based on finding the best
distribution using minimization of $\ell^1$-, $\ell^2$- and
$\ell^{\infty}$-distances and $L^1$-, $L^2$- and
$L^{\infty}$-metrics for GNB and GG distributions, respectively.
This approach, first, allows to obtain parameter estimates without
using grid methods and solving systems of nonlinear equations and,
second, yields not point estimates as the methods of moments or
maximum likelihood do, but the estimate for the density function. In
other words, within this approach the set of decisions is not a
Euclidean space, but a functional space.

\section{The GNB and GG distributions}

It will be assumed that all the random variables are defined on the
same probability space $(\Omega,\,\mathfrak{F},\,{\mathbb P})$.

A random variable having the gamma distribution with shape parameter
$r>0$ and scale parameter $\mu>0$ will be denoted $G_{r,\mu}$,
\begin{equation}
\label{Gamma}
\Pr (G_{r,\mu}<x)=\int_{0}^{x}g(z;r,\mu)dz,\ \ \text{with}\ \
g(x;r,\mu)=\frac{\mu^r}{\Gamma(r)}x^{r-1}e^{-\mu x},\ x\ge0,
\end{equation}
where $\Gamma(r)$ is Euler's gamma-function,
$\Gamma(r)=\int_{0}^{\infty}x^{r-1}e^{-x}dx$, $r>0$.

A GG distribution is the absolutely continuous distribution defined
by the density
\begin{equation}
\label{GG}
g^*(x;r,\gamma,\mu)=\frac{|\gamma|\mu^r}{\Gamma(r)}x^{\gamma
r-1}e^{-\mu x^{\gamma}},\ \ \ \ x\ge0,
\end{equation}
with $\gamma\in\mathbb{R}$, $\mu>0$, $r>0$. The distribution
function corresponding to the density $g^*(x;r,\gamma,\mu)$ can be
denoted $F^*(x;r,\gamma,\mu)$.

The properties of GG distributions were described
in~\cite{Stacy1962,Korolev2013}. A random variable with the
density $g^*(x;r,\gamma,\mu)$ will be denoted
$\overline{G}_{r,\gamma,\mu}$. It can be easily made sure that
\begin{equation}\label{GG1}
\overline{G}_{r,\gamma,\mu}\eqd G_{r,\mu}^{1/\gamma},
\end{equation}
and hence,
\begin{equation}\label{GG2}
(\overline{G}_{r,\gamma,\mu})^{\gamma}\eqd G_{r,\mu}.
\end{equation}

The symbol $\eqd$ in~\eqref{GG1} and~\eqref{GG2} denotes the coincidence of distributions.

A random variable $N_{r,p}$ is said to have the negative binomial
(NB) distribution with parameters $r>0$ (``shape'') and $p\in(0,1)$
(``success probability''), if
\begin{equation}
\label{NB}
{\mathbb P}(N_{r,p}=k)=\frac{\Gamma(r+k)}{k!\Gamma(r)}\cdot p^r(1-p)^k,\
\ \ \ k=0,1,2,...
\end{equation}

Let $r>0$, $\gamma\in\r$ and $\mu>0$. We say that the random
variable $N_{r,\gamma,\mu}$ has the GNB distribution, if
\begin{equation}
\label{GNB}
{\mathbb P}(N_{r,\gamma,\mu}=k)=\frac{1}{k!}\int_{0}^{\infty}e^{-z}z^kg^*(z;r,\gamma,\mu)dz,\
\ \ \ k=0,1,2...,
\end{equation}
and $g^*(z;r,\gamma,\mu)$ is determined by formula~\eqref{GG}.

\section{A functional approach to estimation of the parameters of GNB distributions}

The problem of statistical estimation of the parameters of GNB
distribution (for example, the search for maximum likelihood
estimates) is extremely complicated. To find estimators, the
two-stage grid EM-algorithm for the GNB distribution
$F(x;r,\gamma,\mu)$ can be used. At the first stage, the main part
of the support of the mixing distribution is determined. That is, a
bounded interval is determined such that the probability of a
GG distributed mixing random variable to fall into this interval is
insignificantly less than one. This interval is covered by a finite
grid containing (possibly, a very large number) $K\in\mathbb{N}$ of
known nodes $\lambda_1,...,\lambda_K$. The GNB distribution under
study is approximated by the finite mixture of Poisson
distributions:
\begin{equation}\label{Poismixt}
F(x;r,\gamma,\mu)(x+0)\approx\sum_{j=0}^{[x]}\frac{1}{j!}\sum_{i=1}^Kp_ie^{-\lambda_i}\lambda_i^j,\ \ \ x\in\mathbb{R}.
\end{equation}

In the mixture on the right-hand side of \eqref{Poismixt}, only the
parameters $p_1,...,p_{K}$ are unknown. At the second stage, it
remains to use some standard method for fitting the GG distribution
to the histogram-type data $(\mu_1, p_1),..., (\mu_K, p_K)$,
obtained at the first stage. For example, the parameters $r$,
$\gamma$ and $\mu$ can be determined as the point minimizing the
corresponding chi-square statistic or some special least squares
problem.

However, with a fixed grid, the two-stage method yields only
approximate estimates of the parameters of GG distributions.
Moreover, the accuracy of the approximation depends on the choice of
the grid. The estimates can be consistent in the traditional sense
only if the grid mesh becomes infinitely small as the sample size
infinitely increases in an appropriate way. Moreover, the conditions
unifying the rate of decrease of the grid mesh with the rate of
increase of the sample size that provide the statistical consistency
of the estimators are very cumbersome and practically unverifiable.

In this section we present an alternative methodology based on
finding the best GNB distribution using minimization of $\ell^1$-,
$\ell^2$- and $\ell^{\infty}$-distances (they correspond to the
spaces of sequences whose series are absolutely convergent, the
space of square-summable sequences and the space of bounded
sequences, respectively). Namely, the histogram of the initial data
should be obtained. The integer rule is used as bining algorithm
(due to that the observations in the sample are integer), so bins
are created for each value. Let $N_{b}$ be the number of histogram
bins (with a uniform width that equals $1$), $\bf h$ be the vector
of bar heights ($h_i\in [0,1]$ for all $i=1,\ldots,N_{b}$). The
value of each component $h_i$ is equal to the ratio of a number of
observations in the bin to a total number of observations, the sum
of the bar areas is $1$. So, the bars of empirical distribution can
be approximated by ones of GNB. For finding estimations of unknown
parameters of generalized negative binomial distributions the
following optimization problems should be solved (the density
$g^*(x;r,\gamma,\mu)$ is determined by~\eqref{GG} and the
probability ${\mathbb P}(N_{r,\gamma,\mu}=k)$ is determined
by~\eqref{GNB}).
\begin{itemize}
\item If the target function is based on $\ell^1$-distance:
\begin{equation}
\label{l1}
(r^*,\gamma^*,\mu^*)=\arg\min_{r,\gamma,\mu} \sum\limits_{k=1}^{N_{b}} \left| \frac{1}{k!}\int_{0}^{\infty}e^{-z}z^kg^*(z;r,\gamma,\mu)dz - h_k\right|.
\end{equation}
\item If the target function is based on $\ell^2$-distance:
\begin{equation}
\label{l2}
(r^*,\gamma^*,\mu^*)=\arg\min_{r,\gamma,\mu} \sqrt{\sum\limits_{k=1}^{N_{b}} \left( \frac{1}{k!}\int_{0}^{\infty}e^{-z}z^kg^*(z;r,\gamma,\mu)dz - h_k\right)^2}.
\end{equation}
\item If the target function is based on $\ell^{\infty}$-distance:
\begin{equation}
\label{lInf}
(r^*,\gamma^*,\mu^*)=\arg\min_{r,\gamma,\mu} \max\limits_{k=\overline{1,N_{b}}} \left| \frac{1}{k!}\int_{0}^{\infty}e^{-z}z^kg^*(z;r,\gamma,\mu)dz - h_k\right|.
\end{equation}
\end{itemize}

Formulas~\eqref{l1}--\eqref{lInf} allow to obtain parameter
estimates without using grid methods. It should be noted that this
methodology can also be used for the classical negative binomial
distribution~\eqref{NB} (the ratio~\eqref{NB} should be used in
formulas~\eqref{l1}--\eqref{lInf} instead of~\eqref{GNB}).

\begin{figure} 
\begin{center}
\includegraphics[width=\textwidth]{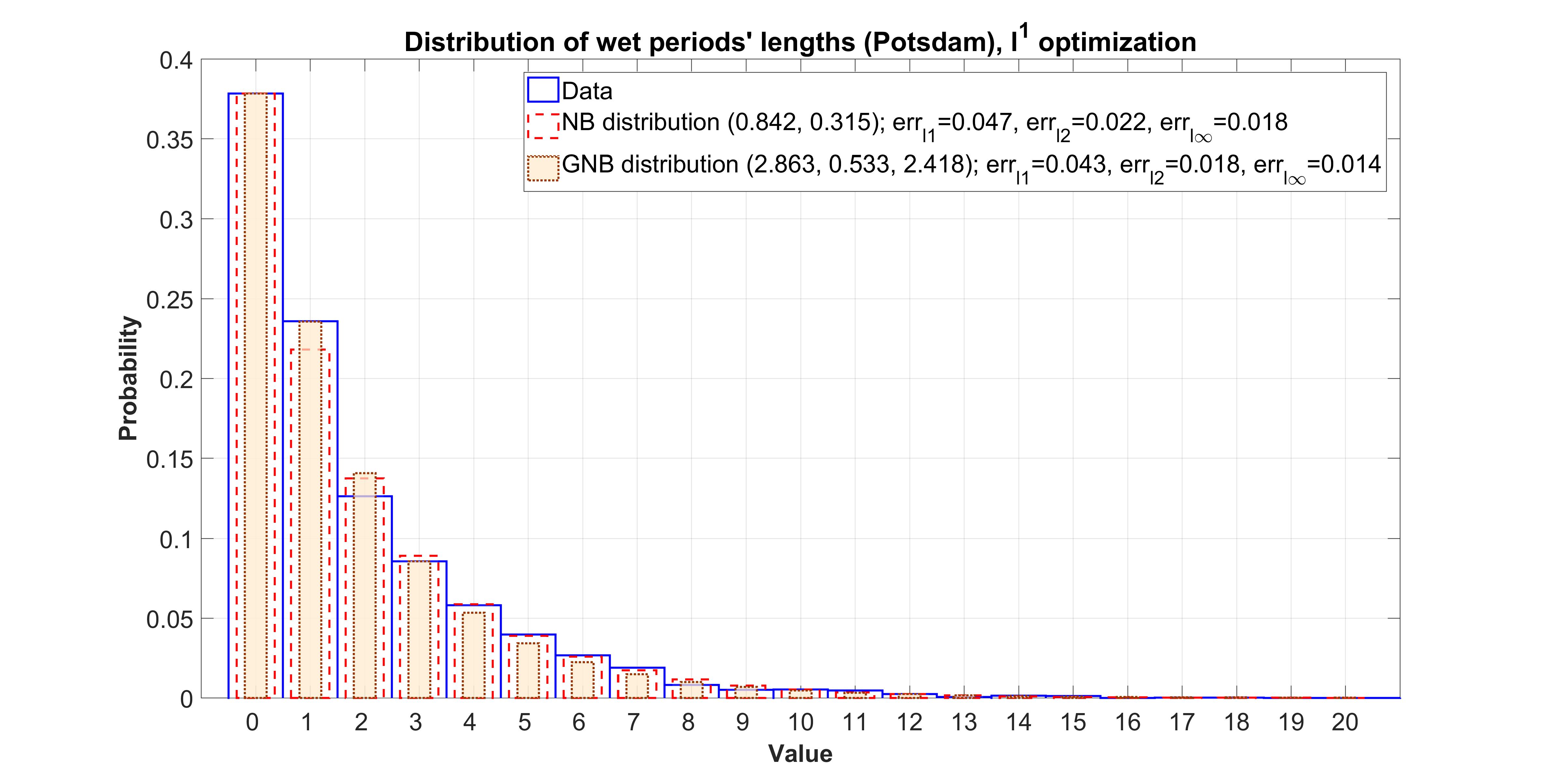}
\end{center}
\caption{\label{FigNB_GNB_l1} Approximation of the initial data distribution by optimization of $\ell^1$-distance.}
\end{figure}

A special~\verb"MATLAB" program is implemented for
finding GNB approximations and plotting figures. The numerical
optimization is based on the simplex search
method~\cite{Lagarias1998}. The functions for estimating the values
of all three unknown parameters of the GNB distribution or two
parameters provided the shape parameter $r$ estimate based on NB
distribution is given are created.

\begin{figure} 
\begin{center}
\includegraphics[width=\textwidth]{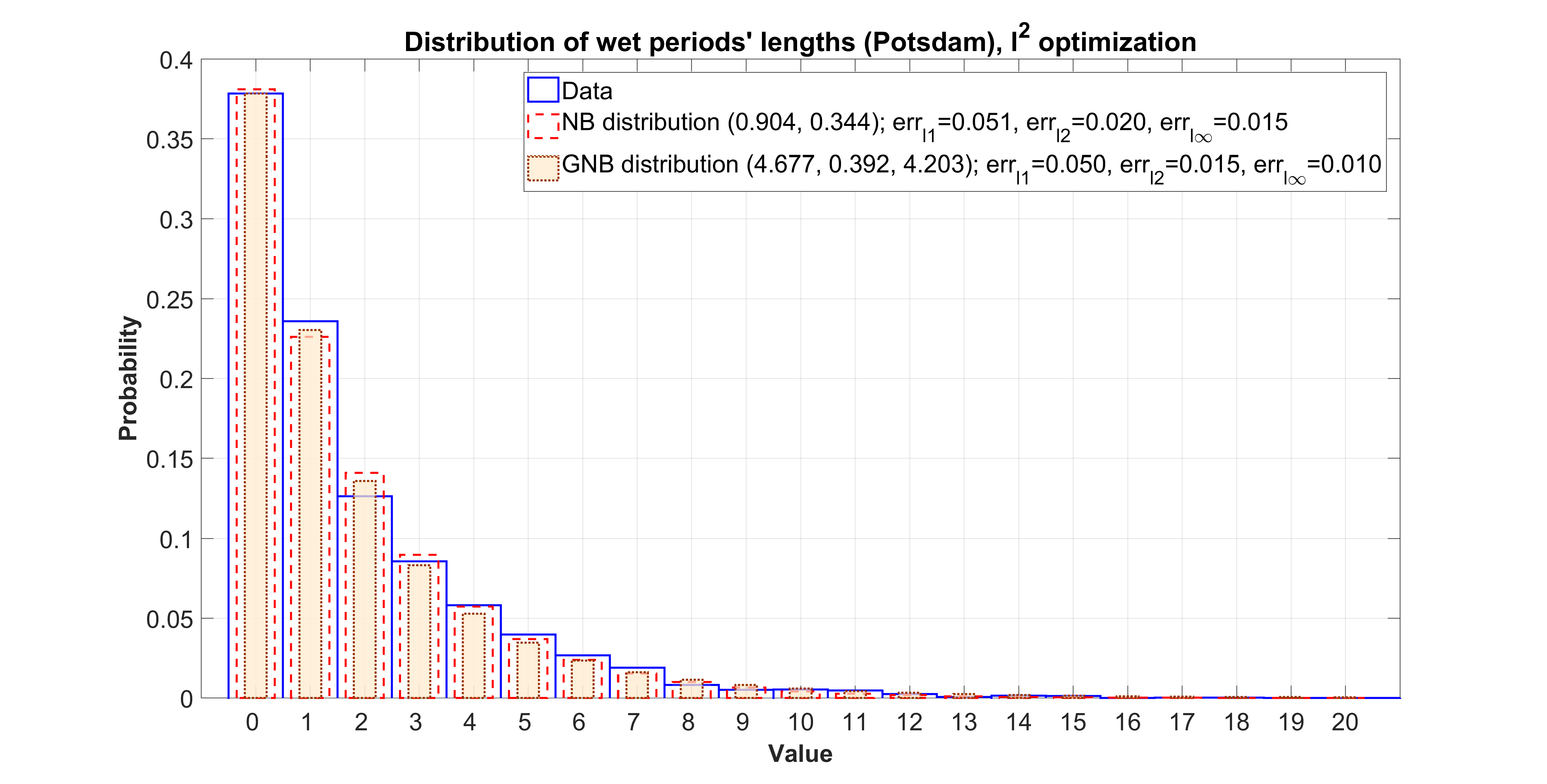}
\end{center}
\caption{\label{FigNB_GNB_l2} Approximation of the initial data distribution by optimization of $\ell^2$-distance.}
\end{figure}

\begin{figure} 
\begin{center}
\includegraphics[width=\textwidth]{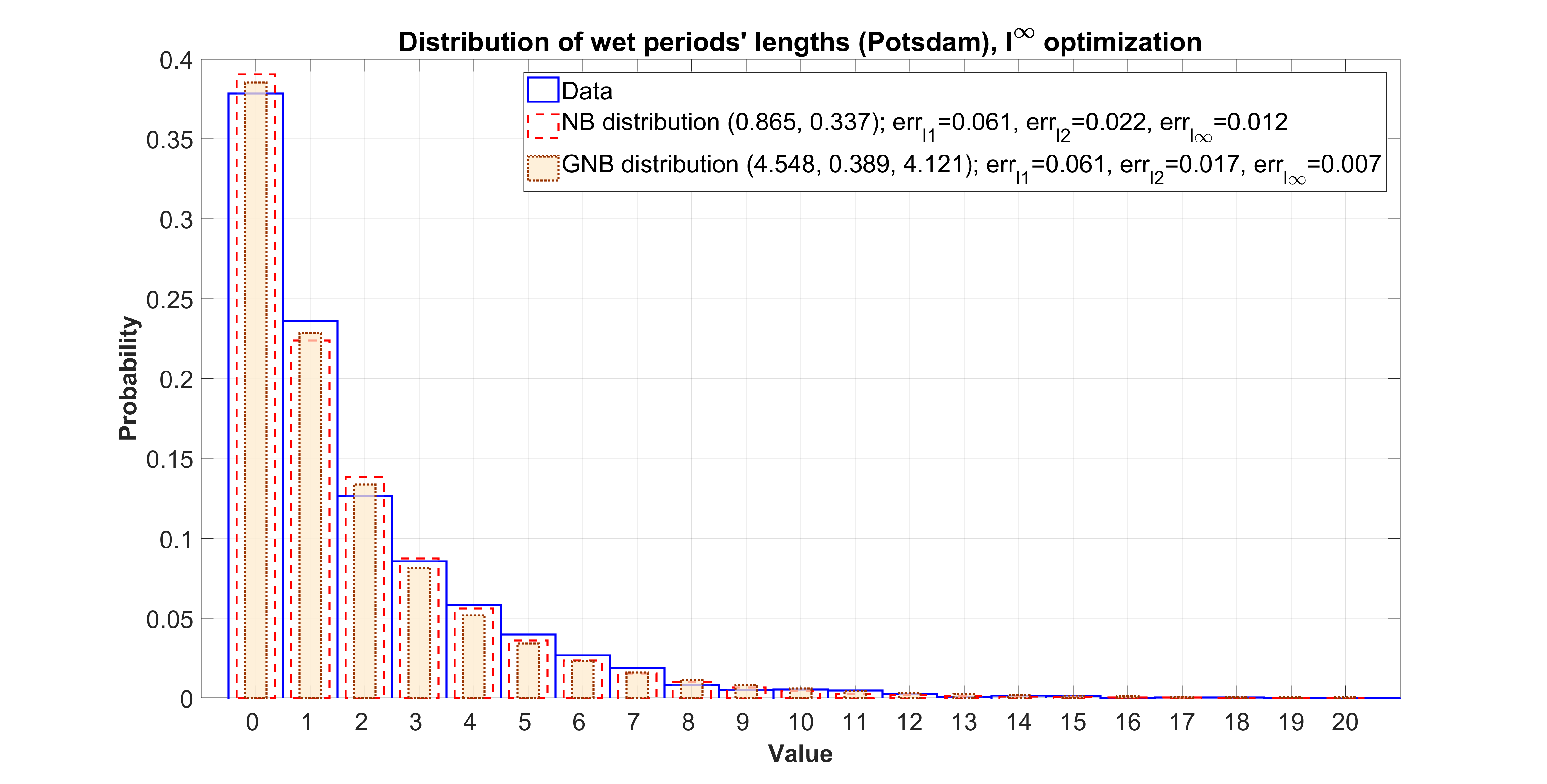}
\end{center}
\caption{\label{FigNB_GNB_l1Inf} Approximation of the initial data distribution by optimization of $\ell^{\infty}$-distance.}
\end{figure}

The histogram and approximating graphs of the NB and GNB
distributions as well as errors in the corresponding metrics are
plotted. The examples of results are shown on
Figs.~\ref{FigNB_GNB_l1}--\ref{FigNB_GNB_l1Inf}. They demonstrate a
high quality of the approximation of the histogram of the initial
data by each type of distributions. Table~\ref{Tab_NB_GNB}
represents approximation errors, the parameters are estimated by
each of the metrics. Obviously, the results for GNB distributions
are better (see the bold marked items).

\begin{table} 
\caption{Approximation errors for negative binomial  and generalized
negative binomial  distributions, test sample.}\label{Tab_NB_GNB}
\centering
\begin{tabular}{|c|c|c|c|}
\hline
{\bf Distribution}&{\bf Error ($\ell^1$)}&{\bf Error ($\ell^2$)}&{\bf Error ($\ell^{\infty}$)} \\
\hline
NB ($\ell^1$-optimization)&$0{,}047$&$0{,}022$&$0{,}018$\\
\hline
GNB ($\ell^1$-optimization)&$\bf 0{,}043$&$0{,}018$&$0{,}014$\\
\hline
NB ($\ell^2$-optimization)&$0{,}051$&$0{,}0195$&$0{,}015$\\
\hline
GNB ($\ell^2$-optimization)&$0{,}05$&$\bf 0{,}015$&$0{,}0097$\\
\hline
NB ($\ell^{\infty}$-optimization)&$0{,}061$&$0{,}022$&$0{,}012$\\
\hline
GNB ($\ell^{\infty}$-optimization)&$0{,}061$&$0{,}017$&$\bf 0{,}007$\\
\hline
\end{tabular}
\end{table}

\section{Recurrence formulas for GNB distributions}

Using formulas~\eqref{GG} and~\eqref{GNB} we can obtain the following results:
\begin{gather*}
\Pr(N_{r,\gamma,\mu}=k)=\frac{|\gamma|\mu^r}{\Gamma(r)k!}\int\limits_{0}^{\infty}
e^{-z-\mu z^\gamma}\, z^{\gamma r+k-1}\,dz= \frac{|\gamma|\mu^r}{\Gamma(r)k!}\int\limits_{0}^{\infty}
e^{-z-\mu z^\gamma} \,d\frac{z^{\gamma r+k}}{\gamma r+k}=\\
=\frac{|\gamma|\mu^r}{\Gamma(r)k!}\times\left[\frac{z^{\gamma r+k}}{\gamma r+k}\,e^{-z-\mu z^\gamma} \bigg|_{0}^{\infty}-
\int\limits_{0}^{\infty}
e^{-z-\mu z^\gamma} \frac{z^{\gamma r+k}}{\gamma r+k} (-1-\mu\gamma z^{\gamma-1})\,dz\right]=\\
=\frac{|\gamma|\mu^r}{\Gamma(r)k!}\times\left[\frac1{\gamma r+k}\int\limits_{0}^{\infty}
e^{-z-\mu z^\gamma} z^{\gamma r+k}\,dz +\frac{\mu\gamma}{\gamma r+k}\int\limits_{0}^{\infty}
e^{-z-\mu z^\gamma} z^{\gamma r+\gamma+k-1}\,dz\right]=\\
=\frac{k+1}{\gamma r+k}\,\Pr(N_{r,\gamma,\mu}=k+1)+\frac{\gamma^2\mu^{r+1}}{(\gamma r+k)\Gamma(r)k!}\int\limits_{0}^{\infty}
e^{-z-\mu z^\gamma} z^{\gamma r+k-1+\gamma}\,dz=\\
=\frac{k+1}{\gamma r+k}\,\Pr(N_{r,\gamma,\mu}=k+1)+\frac{|\gamma|\mu}{\gamma r+k}\Pr(N_{r+1,\,\gamma,\mu}=k).
\end{gather*}

So, the recurrence formulas for GNB distributions can be represented as follows:
\begin{gather}
(\gamma r+k)\,\Pr(N_{r,\gamma,\mu}=k)=(k+1)\,\Pr(N_{r,\gamma,\mu}=k+1)+|\gamma|\mu\,\Pr(N_{r+1,\,\gamma,\mu}=k),\notag\\
\intertext{or}
\Pr(N_{r,\gamma,\mu}=k+1)=\frac{\gamma r+k}{k+1}\,\Pr(N_{r,\gamma,\mu}=k)-\frac{|\gamma|\mu}{k+1}\,\Pr(N_{r+1,\,\gamma,\mu}=k).
\label{Reccurence}
\end{gather}

Unfortunately, the representation~\eqref{Reccurence}
does not significantly simplify the computational process, since in
addition to the value $\Pr(N_{r,\gamma,\mu}=k)$ a value
$\Pr(N_{r+1,\gamma,\mu}=k)$ should be known.

\section{A functional approach to the estimation of the parameters of GG distributions}

In this section we present a methodology based on the search for the
best GG distribution using minimization of $L^1$-, $L^2$- and
$L^{\infty}$-metrics (they correspond to the spaces of functions for
which the $p^{th}$ power of the absolute value is Lebesgue
integrable, where functions that agree almost everywhere are
identified).

The histogram of the initial data should be obtained. The
Freedman--Diaconis rule~\cite{Freedman1981} is used as bining
algorithm due to its suitableness for data with heavy-tailed
distributions. It uses a bin width of
\begin{equation}
\label{Freedman}
2 \,\frac{x_{0{.}75}-x_{0{.}25}}{\sqrt[3]{n}},
\end{equation}
where $x_{0{.}25}$, $x_{0{.}75}$ are $0{.}25$- and
$0{.}75$-quantiles, numerator of fraction~\eqref{Freedman}
represents an interquartile range and $n$ is a sample size.

Let $N_{b}$ be the number of histogram bins, $\bf h$ be the vector
of bar heights ($h_i\in [0,1]$ for all $i=1,\ldots,N_{b}$). The
value of each component $h_i$ is equal to the ratio of the number of
observations in the bin and the total number of observations, the
sum of the bar areas is $1$. Let $\bf b$ be the vector of bin edges.
The bars of empirical distribution should be approximated by
GG distribution.

To find the estimates of unknown parameters of GG distributions the
following optimization problems should be solved (the density
$g^*(x;r,\gamma,\mu)$ is determined by~\eqref{GG}).
\begin{itemize}
\item If the target function is based on $L^1$-metric:
\begin{equation}
\label{L1}
(r^*,\gamma^*,\mu^*)=\arg\min_{r,\gamma,\mu} \sum\limits_{k=1}^{N_{b}-1} \int\limits_{b_k}^{b_{k+1}}\left|g^*(z;r,\gamma,\mu) - h_k\right|\,dz.
\end{equation}
\item If the target function is based on $L^2$-metric:
\begin{equation}
\label{L2}
(r^*,\gamma^*,\mu^*)=\arg\min_{r,\gamma,\mu} \sqrt{\sum\limits_{k=1}^{N_{b}-1} \int\limits_{b_k}^{b_{k+1}}\left(g^*(z;r,\gamma,\mu) - h_k\right)^2}\,dz.
\end{equation}
\item If the target function is based on $L^{\infty}$-metric:
\begin{equation}
\label{LInf}
(r^*,\gamma^*,\mu^*)=\arg\min_{r,\gamma,\mu} \max\limits_{k\in[1,N_b-1]}\int\limits_{b_k}^{b_{k+1}}\left|g^*(z;r,\gamma,\mu) - h_k\right|\,dz.
\end{equation}
\end{itemize}

Formulas~\eqref{L1}--\eqref{LInf} allow to obtain parameter
estimates without using grid methods. It should be noted that this
methodology can also be used for the classical gamma
distribution~\eqref{Gamma}.

\begin{figure} 
\begin{center}
\includegraphics[width=\textwidth]{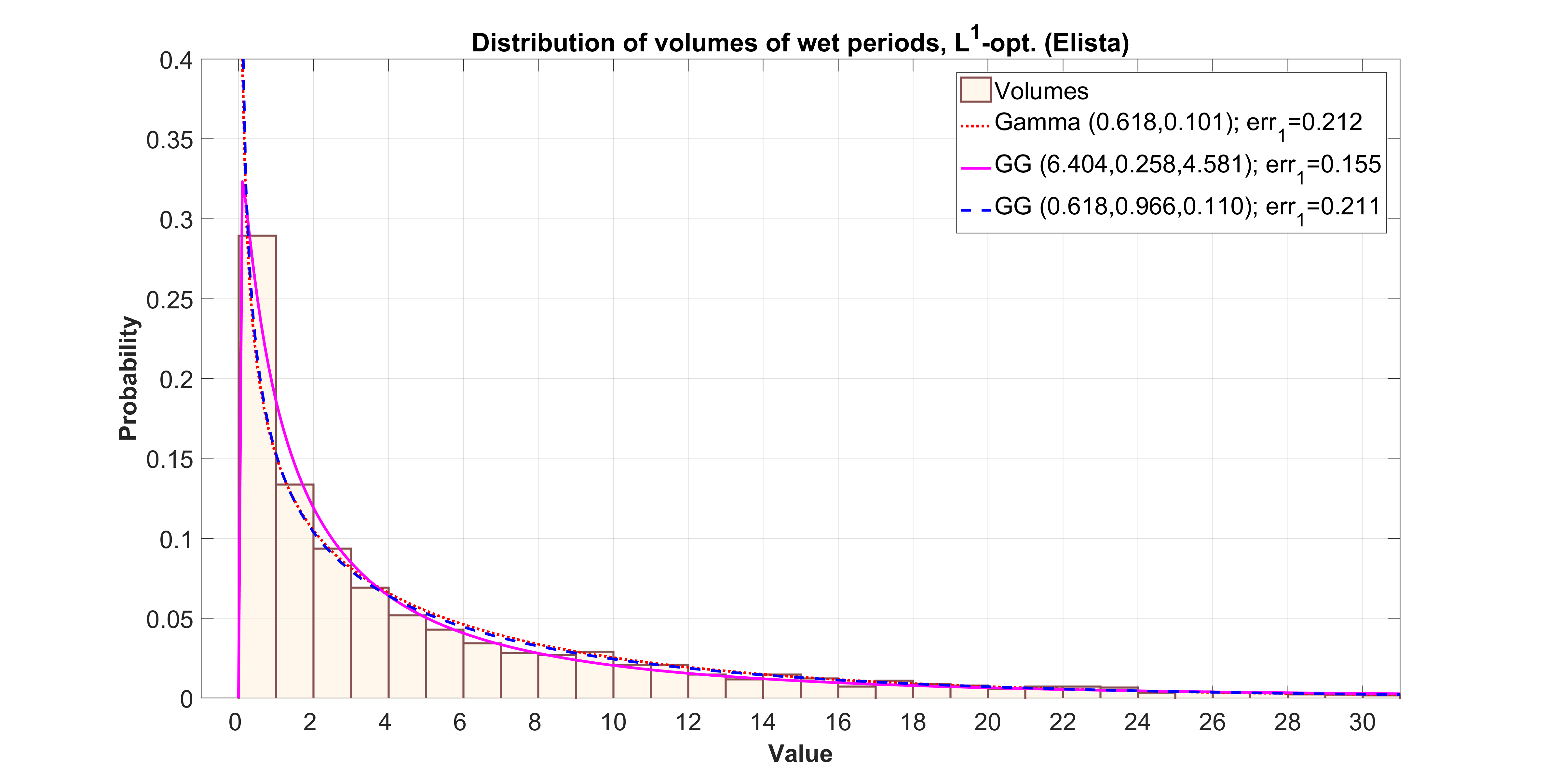}
\end{center}
\caption{\label{FigGG_L1} Approximation of the initial data distribution by optimization of $L^1$-metric.}
\end{figure}

\begin{figure} 
\begin{center}
\includegraphics[width=\textwidth]{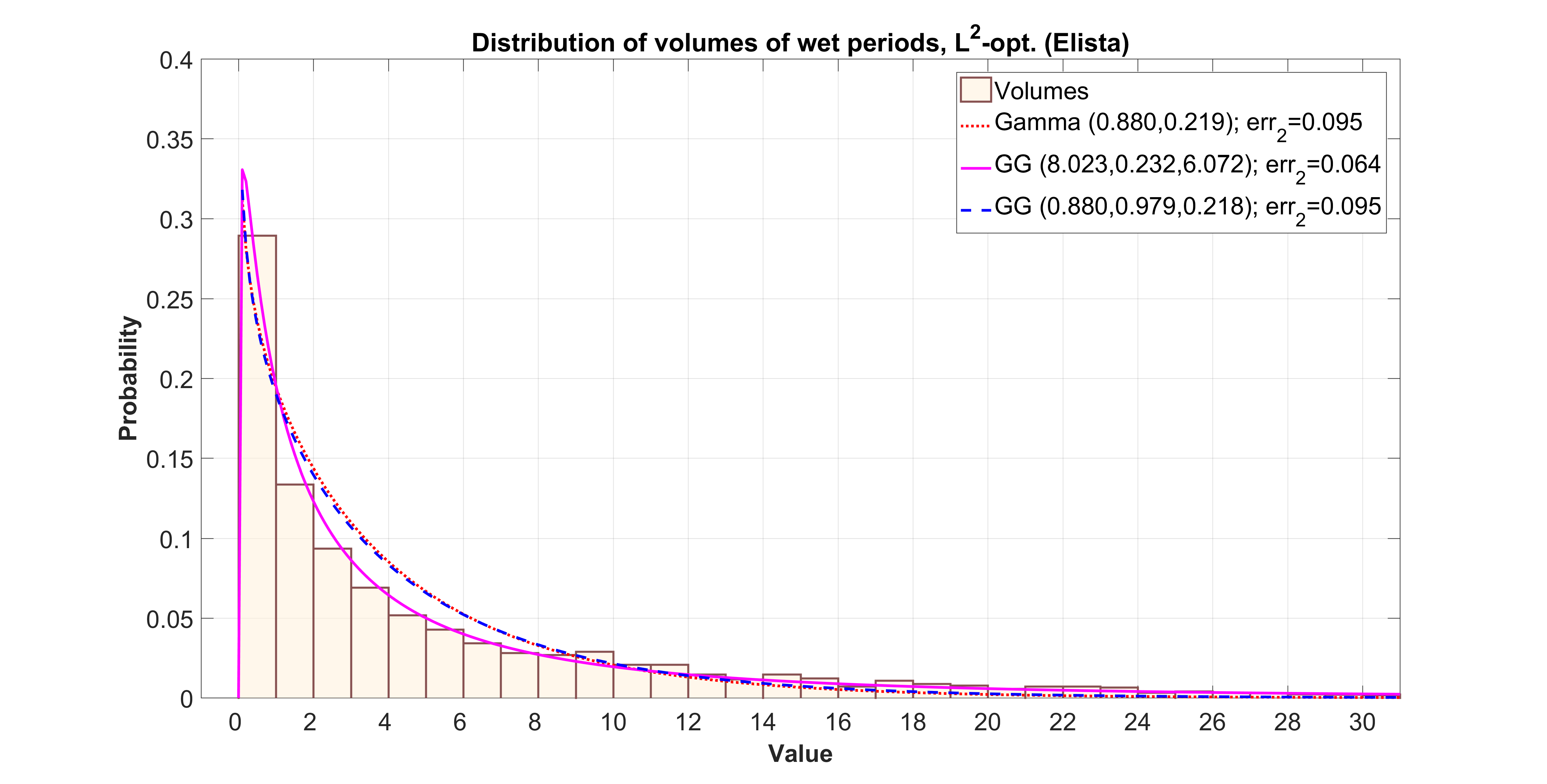}
\end{center}
\caption{\label{FigGG_L2} Approximation of the initial data distribution by optimization of $L^2$-metric.}
\end{figure}

\begin{figure} 
\begin{center}
\includegraphics[width=\textwidth]{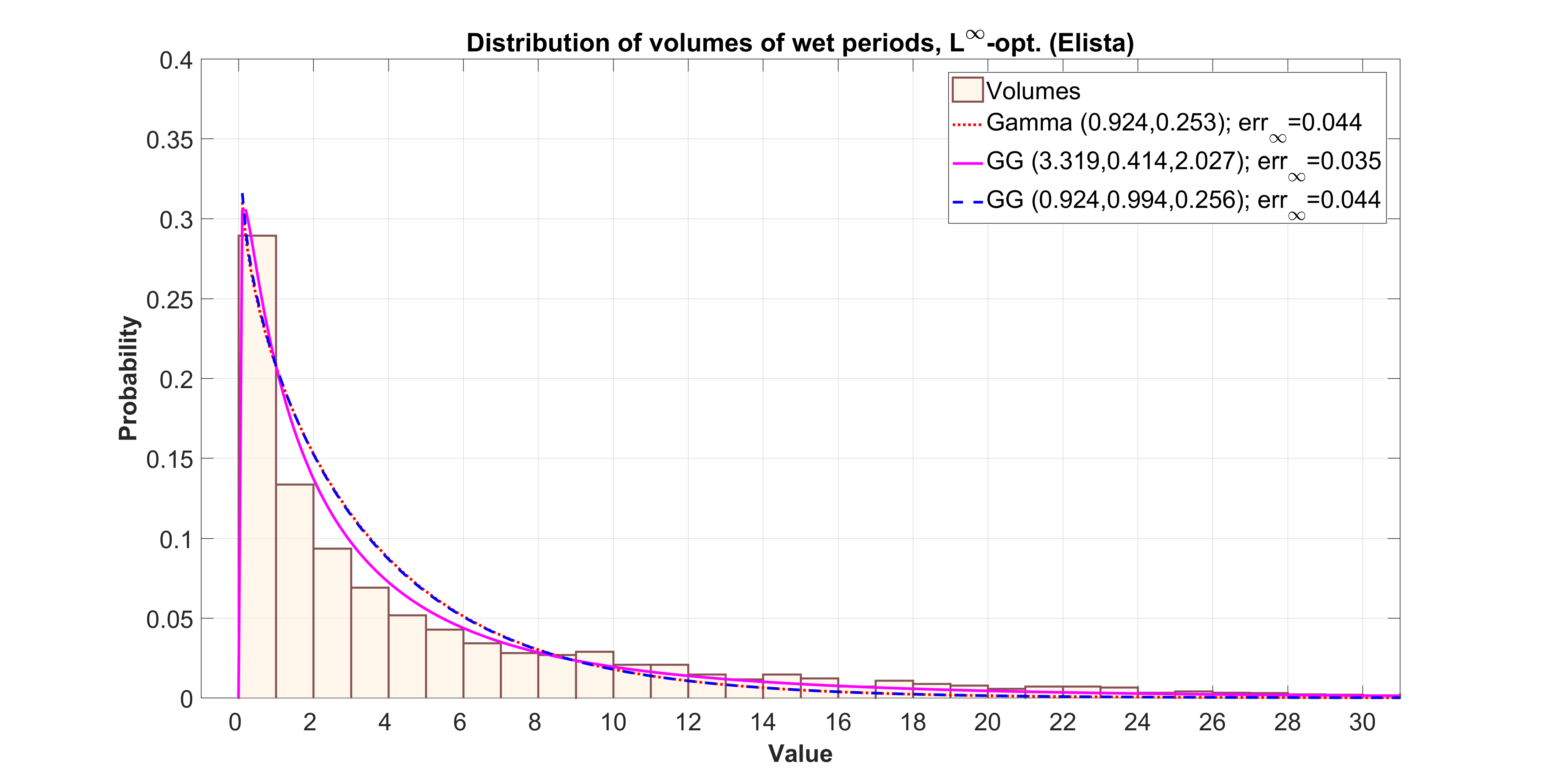}
\end{center}
\caption{\label{FigGG_LInf} Approximation of the initial data distribution by optimization of $L^{\infty}$-metric.}
\end{figure}

A special \verb"MATLAB" program is implemented for finding
GG approximations and plotting figures. The numerical optimization
is based on the simplex search method~\cite{Lagarias1998}. The
functions for estimating values of all three unknown parameters of
GG distribution or two parameters provided the shape parameter $r$
is given based on the gamma distribution model are created. The
histogram and approximating probability density functions of gamma
and GG distributions as well as errors in the corresponding metrics
are plotted. The examples of results are shown on
Figures~\ref{FigGG_L1}--\ref{FigGG_LInf}.

They demonstrate a high quality of the approximation of the histogram of the initial data by each type of distributions. Table~\ref{Tab_Gamma_GG} represents approximation
errors, the parameters are estimated by each of the metrics. Obviously, the results for GG distributions are better (see the bold marked items).

\begin{table} 
\caption{Approximation errors for gamma and generalized gamma distributions, test sample.}\label{Tab_Gamma_GG}
\centering
\begin{tabular}{|c|c|c|c|}
\hline
{\bf Distribution}&{\bf Error ($L^1$)}&{\bf Error ($L^2$)}&{\bf Error ($L^{\infty}$)} \\
\hline
Gamma &$0{,}212$&$0{,}095$&$0{,}044$\\
\hline
GG &$\bf 0{,}155$&$\bf 0{,}064$&$\bf 0{,}035$\\
\hline
GG, fixed $r$&$0{,}211$&$0{,}095$&$0{,}044$\\
\hline
\end{tabular}
\end{table}

\section{Conclusion}

The classical negative binomial distribution was successfully used
as a model for the number of subsequent wet days in precipitation
problems for the data registered in climatically different points
(see, for
example,~\cite{Gorshenin2017a,Gorshenin2017b,Gorshenin2017c}).
It was demonstrated that the fluctuations of the data with very
high confidence fit the negative binomial distribution. Obviously, a
more flexible GNB model could provide even better fit with the
statistical data. Herewith the GG distribution can be
effectively used to model aggregated data (for example, volumes
accumulated over a period) and can be useful for statistical testing
of hypotheses about their extremality.

Moreover, such types of mixed probability models are
quite adequate for information systems (for example, in
insurance~\cite{Grandell1997,Bening2002}, financial
mathematics~\cite{Gorshenin2013a,Korolev2015},
physics~\cite{Korolev2006,Gorshenin2012,Gorshenin2013b}, data
flows~\cite{Gorshenin2013c} and many other fields).
The developed functional methods for the estimation of
the unknown distribution parameters can be implemented as numerical
procedures in the research support system for stochastic data
processing~\cite{Gorshenin2016,Gorshenin2017d} to analyze events in
various information flows.

\section*{Acknowledgments.}

The research is partially supported by the Russian Foundation for
Basic Research (project~17-07-00851) and the RF Presidential
scholarship program (No.~538.2018.5).

\end{document}